\documentclass[final,5p,times,twoside]{elsarticle}

\usepackage{graphicx}

\usepackage{amssymb,amsmath}

\usepackage{dcolumn}  
\usepackage{array}   %
\usepackage{threeparttable}

\newcommand{\gl}{\mathrel{\rlap{\lower1.5pt\hbox{\small \hskip1pt$<$}}
\hspace*{1pt}\raise3pt\hbox{\small $>$}}}

\journal{Physics Letters B}

\begin{document}

\begin{frontmatter}



\title{SU(2) lattice gauge theory in 2+1 dimensions:\\ critical couplings
from twisted boundary conditions and universality}


\author{Sam Edwards and Lorenz von Smekal}

\address{Institut f\"ur Kernphysik, Technische Universit\"at
  Darmstadt, Schlossgartenstr.~9, 64289 Darmstadt, Germany, and\\
 Centre for the Subatomic Structure of Matter (CSSM), School of
   Chemistry \& Physics, University of Adelaide, SA 5005, Australia}

\begin{abstract}
We present a precision determination of the critical coupling
$\beta_c$ for the deconfinement transition in pure SU(2) gauge
theory in $2+1$ dimensions. This is possible from universality, 
by intersecting the center vortex free energy as a function of the
lattice coupling $\beta $ with the exactly known value of the
interface free energy in the 2D Ising model at criticality. Results
for lattices with different numbers of sites $N_t$ along the Euclidean time
direction are used to determine how $\beta $ varies with temperature
for a given $N_t$ around the deconfinement transition. 
\end{abstract}


\end{frontmatter}


\section{Introduction}
\label{intro}

It is widely accepted today that the deconfinement transition in
pure SU($N$) gauge theories at finite temperature is driven by the
dynamics of center vortices \cite{Greensite:2003bk}.  In the vortex
picture of confinement, Wilson loops acquire a disordering phase
factor from every vortex that they link with. The area law for
timelike Wilson loops in the pure gauge theory comes from the
percolation of spacelike vortex sheets in the confined phase. Their
free energies have been measured over the deconfinement phase
transition at finite temperature in the $3+1$ dimensional pure SU(2)
gauge theory from ratios of partition functions with 't Hooft's
twisted boundary conditions in temporal planes, forcing odd numbers of
$Z_2$ center vortices through those planes, over the periodic ensemble
with even numbers \cite{deForcrand:2001nd,vonSmekal:2002ps}.   
A Kramers-Wannier duality is then observed by comparing the
behaviour of these center vortices with that of 't Hooft's electric
fluxes which yield the free energies of static charges in
a well-defined (UV-regular) way~\cite{deForcrand:2001dp}, with
boundary conditions to mimic the presence of 'mirror' (anti)charges in
neighbouring volumes. This duality follows that between the Wilson
loops of the 3-dimensional $Z_2$-gauge theory and the 3D-Ising spins,
reflecting the universality of the center symmetry breaking transition. 
Here we study the vortex free energies of pure SU(2) in $2+1$
dimensions over the deconfinement transition because the relevant
interface free energies of its universal partner, the self-dual Ising
model in 2 dimensions, are all known analytically. Moreover, the
vortex free energies in $2+1$ dimensions are much cheaper to
simulate and discretisation effects vanish more rapidly than in
$3+1$ dimensions. Together these reasons allow for numerical studies 
of much higher precision. 

In Sections 2 and 3 we briefly outline the basic concepts and our
numerical procedure. Our results are presented in Sec. 4. These
include the precise determination of the critical coupling for various
lattices with up to $N_t=9$ sites in the Euclidean time direction, an
analysis of the finite volume corrections and a brief comparison with
the corrections to scaling in the 2D Ising model. We then determine
how the critical coupling depends on $N_t$, including $1/N_t$
corrections, and use this result to derive how the lattice coupling $\beta $
varies with temperature for a given fixed $N_t$ around the
deconfinement transition. This is needed, for example, for a detailed
finite-size scaling analysis of the vortex free energies in 2+1
dimensional SU(2) \cite{SamEinPrep}.

\section{Concepts and Methods}

For pure SU(2) gauge theory, 't Hooft's twisted boundary conditions
fix the total number of $Z_{2}$ vortices modulo 2 through each plane
of a finite box \cite{'tHooft:1979uj}. Twist in a plane corresponds
to an ensemble with an odd number of $Z_{2}$ vortices through that
plane.

In a $L^{2}\times1/T$ Euclidean box, we can distinguish between two
types of twist. Magnetic twist is defined in the purely spatial plane
and forces vortices that run along the temporal direction. They may
spread independently of the temperature $T$ and play no role in the
deconfinement transition: their free energy is expected to vanish in
the thermodynamic limit at all temperatures which has been
demonstrated explicitly in $3+1$ dimensions \cite{vonSmekal:2002gg}.   
On the other hand, vortices from twist in the two temporal planes are
squeezed more and more as $T$ is increased. They may no longer spread
arbitrarily and this is what drives the phase transition.

In this paper we're interested in configurations with a twist in one
of the temporal planes. If we denote the partition function of this
ensemble by $Z_{tw}(L,T)$, its free energy per $T$ is defined via
the ratio 
\begin{equation}
Z_{tw}(L,T)/Z_{0}(L,T)=e^{-F_{tw}(L,T)}\label{eq:part ratio}
\end{equation}
where $Z_{0}$ is the partition function of the periodic ensemble. 

As we approach the thermodynamic limit, $Z_{tw}/Z_{0}$ converges
to a non-trivial fixed point at the critical temperature. 
It can therefore be used as a phenomenological coupling
in the same manner as the Binder cumulant. Typically, one uses the
pairwise intersections of Binder cumulant curves from different lattice
sizes to estimate the critical coupling or temperature
\cite{Binder:1981sa,Binder:2001ha}. 
Using a ratio of partition functions in this way was proposed by
Hasenbusch for the 3D Ising model \cite{Hasenbusch:1993wn}.

Here we can go a step further. Since SU(2) in 2+1 dimensions is in
the same universality class as the 2D Ising model, the the value of
$Z_{tw}/Z_{0}$ at the fixed point is exactly known. A single temporal
twist in SU(2) corresponds to a single anti-periodic direction on
a $N\times N$ Ising lattice. This forces
an odd number of spin interfaces perpendicular to the anti periodic
direction. The corresponding ratio of partition functions $Z_{ap}/Z_{pp}$
in the Ising model gives the free energy of the system in the same
way as equation (\ref{eq:part ratio}). In both cases, configurations
of minimum action/energy dominate the partition function. Thus, in
the thermodynamic limit, the free energy of a single spacelike vortex
in SU(2) is identified with the free energy of a single interface
in the 2D square Ising model at the respective critical points. 
This universal value is given by \cite{PhysRevB.38.565} 
\begin{equation}
\lim_{N\rightarrow\infty}Z_{ap}(T_{c})/Z_{pp}(T_{c})=1/(1+2^{3/4}).
\label{eq:universal ratio}
\end{equation}

\begin{figure}
\includegraphics[width=1\columnwidth]{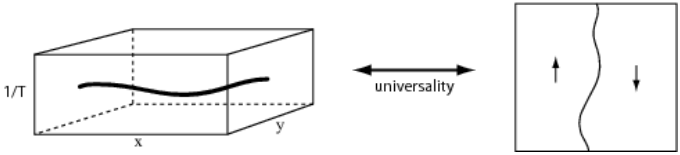}
\caption{Universality}
\end{figure}

For SU(2) on the lattice, $1/T=N_{t}a$ where $N_{t}$ is the number
of sites in the time direction and the lattice spacing $a\equiv a(\beta)$
depends on the coupling. So the critical temperature $T_{c}$ corresponds
to a critical lattice coupling $\beta_{c,\infty}$ for each $N_{t}$.
The subscript $\infty$ reminds us that we only have a strict phase
transition and hence critical coupling in the limit of infinite spatial
volume. Still, the intersection of $Z_{tw}/Z_{0}$ with the universal
value (\ref{eq:universal ratio}) gives a reliable estimate of
$\beta_{c,\infty}(N_{t})$ 
provided that the spatial length $L=N_{s}a$ is large enough. This
will be  more precise in general than the estimates obtained via pairwise
intersections.\footnote{We thank M. Hasenbusch for pointing out
  Ref.~\cite{Caselle:1995wn}, where 
a similar idea was applied to the $Z_{2}$ gauge theory in 2+1 dimensions.}

We assume a finite size scaling (FSS) behaviour of the form 
\begin{equation}
Z_{tw}/Z_{0}=1/(1+2^{3/4})+b(\beta-\beta_{c})N_{s}^{1/\nu}+cN_{s}^{-\omega}+\dots
\; , 
\label{eq:ratio fss}
\end{equation} 
where $\omega$ is a correction to scaling exponent that should be
approximately independent of $N_{t}$, and $\nu=1$ is exactly known
from the 2D Ising model. Furthermore, $\beta_c \equiv
\beta_{c,\infty}(N_{t})$, and we define a 'pseudo-critical
coupling' in a finite volume, $\beta_c(N_t,N_s)$,  by the requirement that the 
corrections to the universal value in (\ref{eq:ratio fss}) 
vanish. These estimates then converge to the infinite volume critical coupling
$\beta_{c,\infty}(N_{t})$ as 
\begin{eqnarray}
\beta_{c}(N_{t},N_{s}) & = &
\beta_{c,\infty}(N_{t}) - d(N_t) \, N_{s}^{-(\omega+1/\nu)}+\dots
\label{eq:beta scaling}
\end{eqnarray}
Here, the coeffiecient $d = c/b$ is an $N_{t}$
dependent fit parameter. 

For notational simplicity we will hereafter drop the subscript
$\infty$ on the critical coupling. Unless $N_{s}$ is explicitly
mentioned, $\beta_{c}$ refers to the infinite volume limit. %
We have used a variety of spatial lattice sizes and the above scaling
formula (\ref{eq:ratio fss}) to first find $\beta_c(N_t,N_s)$ for $N_{t} =
4,\, 5,\, 6,\, 7, \, 8 $ and $ 9$ with high precision, in order to
then determine the corresponding $\beta_c(N_{t})$  from fitting
the volume dependence of $\beta_c(N_t,N_s) $ by (\ref{eq:beta scaling}).

\begin{figure}

\leftline{\hskip -1cm\includegraphics[width=1.14\columnwidth]{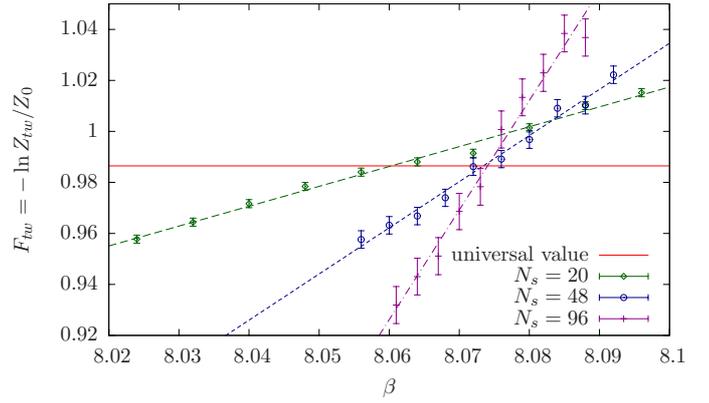}}
\caption{\label{fig:Selected-fits}Fitted curves for $N_{t}=5$, selected for
clarity.}
\end{figure}

\section{Numerical recipe}

Twist was implemented in the usual way \cite{Kajantie:1990bu}.
We kept periodic boundary conditions but flipped the coupling
$\beta\rightarrow-\beta$ 
of a stack of plaquettes perpendicular to the plane of the twist.
In other words, we introduced a $Z_{2}$ Dirac string that is closed
by lattice periodicity. The result is a transformation of the usual
Wilson action.

In practice, the overlap of $Z_{tw}$ and $Z_{0}$ is poor. To overcome
this we interpolated in the number of flipped plaquettes using the
snake algorithm of Ref.~\cite{deForcrand:2000fi}.
This was combined with the other variance reduction tricks therein.

For each combination of $N_{t}$ and $N_{s}$ we performed simulations
for $\sim10$ values of $\beta$ around the intersection of $Z_{tw}/Z_{0}$
with the exact Ising value (\ref{eq:universal ratio}). Random errors
were estimated via the boostrap method. Since SU(2) lattice gauge
theory is less computationally expensive in 2+1 dimensions than in
3+1 dimensions, we were able to perform a very large number of measurements.
$1-30$ million configurations were used for each $\beta$, depending
on the lattice size.

It turns out that the free energy $F_{tw}(\beta)=-\ln Z_{tw}/Z_{0}$ has
less curvature than $Z_{tw}/Z_{0}$ near the critical point, so it's
a better candidate for linear approximation. Therefore we performed least
squares fits, with parameters $f_1$ and $\beta_c$, of the form 
\begin{equation}
F_{tw}(\beta) = f_1 \, (\beta-\beta_{c})+\ln(1+2^{3/4})
\label{eq:linear fit}
\end{equation}
 to obtain estimates $\beta_c = \beta_{c}(N_{t},N_{s})$ for the
 critical lattice coupling. In each case the reduced $\chi^{2}$ was
 $\sim1$, which justifies the 
linear ansatz (\ref{eq:linear fit}). See Fig. \ref{fig:Selected-fits}
for some representative fits. %

\begin{figure}[t]
\includegraphics[width=1\columnwidth]{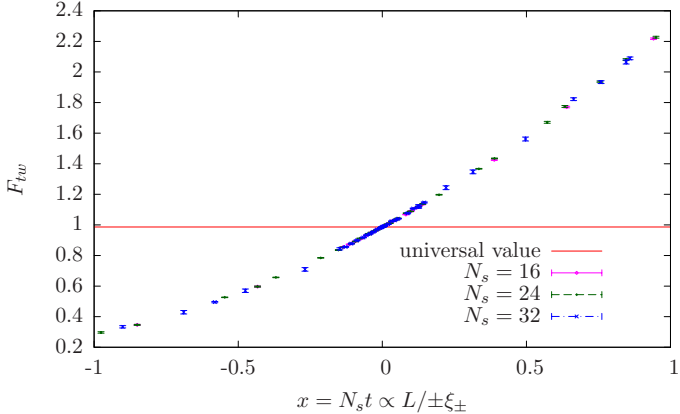}

\caption{FSS data collapse for $16^{2}\times4$, $24^{2}\times4$ and $32^{2}\times4$. \label{fig:FSS-data-collapse}}

\end{figure}

Note, however, that a small reduced $\chi^{2}$ does not exclude the
existence of significant systematic errors in
$\beta_{c}(N_{t},N_{s})$. 
On finite lattices, $F_{tw}(\beta)$ has positive curvature near the critical
coupling, so $\beta_{c}$ tends to be underestimated. To control
this, we carefully chose the size of our fitting windows. For each
$N_{t}$, we performed precise measurements of $F_{tw}$ in a quadratic
fitting window for one or more of our smallest lattices. By writing
$F_{tw}$ as a function of the finite size scaling variable
\begin{equation}
x = N_s\, t \propto  \pm L/{\xi_\pm }\, ,
\end{equation}
where $t=T/T_c-1$ is the reduced temperature and $\xi=\xi_\pm^0 |t|^{-\nu}$
are the correlation lengths for $T \gl T_c $ with $\nu=1$ for the 2D
Ising model, we were able to translate
this data to larger $N_{s}$. See Fig. \ref{fig:FSS-data-collapse} for
an example of FSS data collapse in a large window for several lattice
sizes.\footnote{We will present a more extensive treatment of the FSS
  of $F_{tw}$ in a forthcoming paper.}
The relationship between temperature and coupling is described in Section \ref{sub:betac vs T}. It requires a paramerisation of $\beta_{c}$ vs $N_{t}=1/T_{c}a_{c}$, which we roughly obtained using literature
values of the critical coupling and our own preliminary results.
The translated data was used to estimate the slope and curvature
of $F_{tw}(\beta)$ near the critical point for large lattices
without performing additional simulations. We then adjusted the linear
fitting windows such that the systematic errors in $\beta_{c}(N_{t},N_{s})$
should  be less than one quarter of the quoted random error for each
extrapolated $\beta_{c}(N_{t}).$

\section{Results}

\subsection{Determining $\beta_{c}(N_{t})$}

\begin{figure}[b]
\leftline{\hspace{-1.3cm} 
\includegraphics[width=1.19\columnwidth]{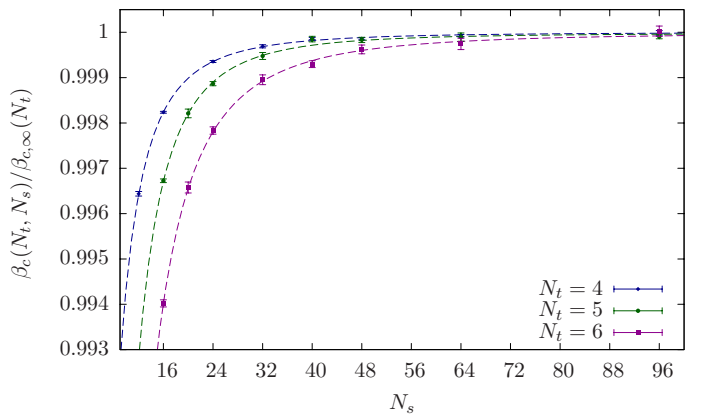}}

\caption{Critical beta estimates
  vs $N_{s}$  relative to their infinite volume limits for $N_{t}=4,\,
  5$ and $6$, and the corresponding fits to (\ref{eq:beta
  scaling}). \label{fig:Critical-beta-estimate-nt456}}

\end{figure}

We have obtained estimates of $\beta_{c}(N_{t})$ for every $N_{t}$
between 4 and 9. In each case we used 8 spatial lattice sizes, keeping
an aspect ratio of approximately 3:1 for the smallest lattice and
using a maximum lattice size of $96^{2}\times N_{t}$. See 
Fig.~\ref{fig:Critical-beta-estimate-nt456}
for plots of $\beta_{c}(N_{t},N_{s})$ vs $N_{s}$ for $N_{t}=4$,
5 and 6. The data was fitted according to the equation (\ref{eq:beta scaling})
with all three parameters free. The plots for $N_{t}=$
7, 8 and 9 look very similar. Note the rapid convergence to
the infinite volume values, which Hasenbusch also observed for the
pairwise intersection of $Z_{ap}/Z_{pp}$ curves in the 3D Ising model
\cite{Hasenbusch:1993wn}. He found much more rapid convergence than
for the intersections of Binder cumulants.

\begin{figure}
\includegraphics[width=1\columnwidth]{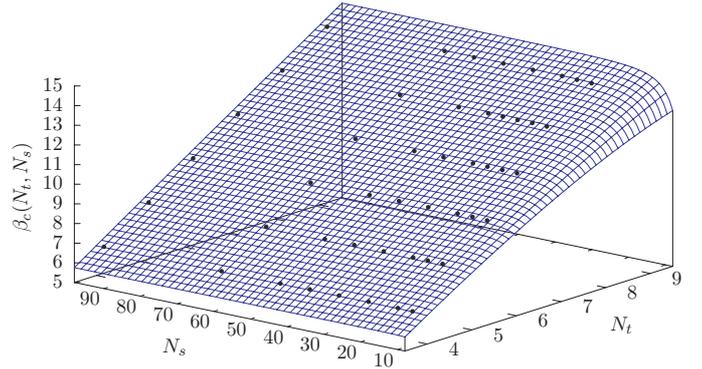}

\caption{All values for $\beta_c(N_t,N_s)$ obtained in this
  work. The error bars are generally smaller than the symbols, and the surface
  was rendered from the fits to (\ref{eq:beta scaling}) with $d(N_t) =
  0.134\, N_t^{\omega+2} - 0.51 \, N_t^{\omega}$ according to
  (\ref{eq:d_fit_2}) with $d_1=0$.
   \label{fig:Critical-beta-estimate-all}}

\end{figure}

\begin{table*}[ht]
 
\begin{threeparttable}

\begin{tabular}{@{\hskip 2pt}c@{\hskip 8pt}p{2cm}cccccccp{1.5cm}}
$N_{t}$ &  $N_s$ used & $\beta_{c}(N_{t},64)$ & 
$\beta_{c}(N_{t},96)$ & $\beta_{c,\infty}(N_{t})$  & $\omega$ (fit)  & 
$\chi^{2}$/dof.  &  $\beta_{c,\infty}(N_{t})\big|_{\omega=1.61}$ &
  $\chi^{2}$/dof. & Lit. values\\[4pt] 
\hline\\[-11pt] 
\hline\\[-8pt] 
4 & 12, 16, 24, 32, 40, 48, 64, 96  &  6.53611(19) & 6.53648(37) 
  & 6.53661(13) & 1.47(6) & 0.98 & 6.53640(10)& 1.62 &

6.483(26)\tnote{$\dagger$}\hfill\break  
6.52(3)\tnote{$\ddagger$}\hfill\break 6.588(25)\tnote{$\S$} \\[4pt] 
\hline\\[-8pt]  
5 & 16, 20, 24, 32, 40, 48, 64, 96  & 8.07392(74) & 8.07402(39)
  & 8.07463(38)  & 1.73(15)  & 1.35  & 8.07488(26) & 1.29
  & 8.143(57)\tnote{$\dagger$} \\[1pt]
\hline\\[-8pt]  
6 & 16, 20, 24, 32, 40, 48, 64, 96 &  9.6002(12) &  9.6029(11)
  & 9.60265(49)  & 1.48(7)  & 0.37  & 9.60185(33) & 0.60
  & 9.55(4)\tnote{$\ddagger$}\\[1pt]
\hline\\[-8pt]  
7 & 20, 24, 28, 32, 40, 48, 64, 96 & 11.1164(15) &  11.1181(36)
  & 11.1194(29) & 1.38(43) & 1.74554 & 11.1181(13) & 1.54 & -- \\[1pt]
\hline\\[-8pt]  
8 & 24, 28, 32, 36, 40, 48, 64, 96  & 12.6301(32) & 12.6342(53) 
  & 12.6348(40) & 1.66(52) & 0.89 & 12.6348(19) & 0.89 & -- \\[1pt]
\hline\\[-8pt]  
9 & 24, 28, 32, 40, 48, 56, 64, 96 & 14.1488(94) &  14.131(11) 
  & 14.1418(68) & 1.96(79) & 0.60 & 14.1446(39) & 0.52 & -- 
\end{tabular}


\end{threeparttable}

\caption{Summary of results from the $N_s^2 \times N_t$ lattices
  specified in columns 1 and 2 with the critical couplings for the largest
  two $N_s$ given explicitly in columns 3 and 4; the infinite volume
  extrapolations from fits to Eq.~(\ref{eq:beta scaling}) are shown in
  column 5 with the resulting exponents $\omega$ and 
  $\chi^2$/dof. in columns 6 and 7. Columns 8 and 9 show the 
  same extrapolations when using the global average $\omega = 1.61(9)$
  in all fits. 
  Literature values are quoted for comparison from Refs.\ $^\dagger$
  \cite{Liddle:2008kk}, $^\ddagger$ \cite{Engels:1996dz} and $^\S$
  \cite{Teper:1993gp}.} \label{tab:Critical-beta-values}

\end{table*}

We summarize our results in Table \ref{tab:Critical-beta-values}.
For each $N_{t}$ we include the estimates $\beta_{c}(N_{t},N_{s})$
from our two largest lattices as well as the fitted values of
$\beta_{c}(N_{t})$. 
It's clear from the reduced $\chi^{2}$s that the data is very well
described by the FSS ansatz (\ref{eq:beta scaling}). 

For $N_{t}=$ 4, 5 and 6 we were able to surpass the precision of
current literature values of the critical couplings by two orders
of magnitude. These lattices also gave us good precision for the correction
to scaling exponent $\omega$. Our results for $N_{t}=$ 7, 8, 9 are
somewhat less precise, especially the fitted values for $\omega$.
This is because the lattices used had smaller aspect ratios and we
collected fewer statistics. Still, $\omega$ is consistent between
each of the fits. In all, we obtain a weighted average for $\omega$ of 1.61(9),
which is also consistent with the value of 1.64 obtained by Engels
et al.~\cite{Engels:1985tz} in early study of Polyakov loop averages. 

For reference, we have included extrapolations of the critical coupling
with $\omega$ fixed at 1.61. Due to correlations, the quoted errors for
$\beta_{c}(N_{t})|_{\omega=1.61}$ should be taken with a grain of
salt. Nevertheless, fixing $\omega$ may lead to more accurate values
of $\beta_c(N_{t})$ for our large $N_{t}$ results.

\begin{figure}
\rightline{\hskip .2cm
\includegraphics[width=1\columnwidth]{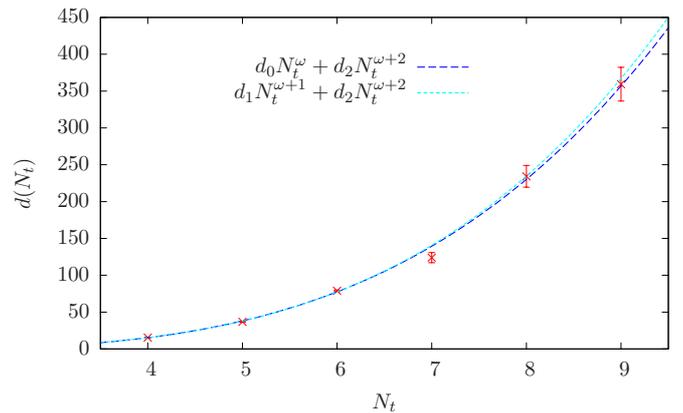}}

\caption{The parameter $d(N_t)$ from the fits of $\beta_c(N_t,N_s) $
  to Eq.~(\ref{eq:beta scaling}) together with 2 parameter fits of its
  $N_t$ dependence via the form in Eq.~(\ref{eq:d_fit_2}) with $d_1 =
  0$ (long-dashed) and $d_0 = 0$ (short-dashed),
  respectively. \label{fig:D_Nt}}

\end{figure}

We furthermore obtain the fit parameter $d(N_t)$ in (\ref{eq:beta scaling})
for each of the six $N_t$ values. 
This in turn allows us to fit its $N_t$ dependence. Using a two parameter form 
$d_\mathrm{fit}(N_t) =  d_\gamma \,  N_t^{\gamma_\mathrm{fit}} $  with a single
effective exponent $\gamma_\mathrm{fit} $, a fairly good description is obtained
with $\gamma_\mathrm{fit} = 3.97(7)$ (and $d_\gamma = 0.06(1)$).  This fit works
best for the smaller $N_t$ but it deteriorates somewhat towards $N_t = 9$.
The $N_t $ dependence of $d(N_t) $ might well be determined by several
competing terms with nearby exponents and is thus difficult to extract
reliably from the data. Alternatively, using 3 parameter fits of the
form 
\begin{equation}
  \label{eq:d_fit_1}
  d_\mathrm{fit}(N_t) = d_\delta\, N_t^\delta   + d_\gamma \,
  N_t^{\gamma_\mathrm{fit}}\, ,
\end{equation}
we obtain $\gamma_\mathrm{fit}  = 3.8(2)$ for $\delta = 0$, 
$\gamma_\mathrm{fit}  = 3.7(4)$ for $\delta = \omega$, or 
$\gamma_\mathrm{fit}  = 3.5(6)$ for $\delta = \omega +1 $, 
for example, where we have used $\omega = 1.61$ corresponding to our
global fit for the correction to scaling exponent $\omega$. 
Because $\omega + 2 = 3.61 $, this suggests that one might also try a
form
\begin{equation} 
 \label{eq:d_fit_2}
  d_\mathrm{fit}(N_t) = d_0 \, N_t^\omega   + d_1 \,
  N_t^{\omega+1} + d_2\, N_t^{\omega+2}\, ,
\end{equation}
with $\omega $ fixed at 1.61. This form is particularly interesting
because it will allow us below to parametrise the finite-size corrections
entirely in terms of the aspect ratio in the form
$N_t/N_s$. Unfortunately, the
coefficients $d_0$ and $d_1$ of the two subleading powers in $\omega $
are much too correlated to determine all the 3 parameters in this fit
reliably from the available data. But we can get good fits with either
$d_1 = 0$ or $d_0= 0$, see Fig. \ref{fig:D_Nt}. Keeping the resulting
two parameters fixed afterwards, a further fit to the remaining one
($d_1$ or $d_0$) then reproduces a zero result with very high accuracy
({\it i.e.}, values smaller than $ 10^{-4}$ in either case).

It's interesting to check how much was gained by using the exact universal
value (\ref{eq:universal ratio}) instead of pairwise intersections.
To this end we performed the pairwise intersection method for $N_{t}=4$.
See Table \ref{tab:Comparison-of-couplings} for the results. The
left side of the table shows the intersection coupling for lattices with
$N_{s}$ and $N_{s}^{'}=N_{s}/2$. On the right hand side are comparative results from intersections with the universal Ising
reference line.

It follows from Eq. (\ref{eq:ratio fss}) that the pairwise intersection
points should scale like Eq. (\ref{eq:beta scaling}), except with
$-d\rightarrow 2(2^{\omega}-1)d$. As such, we've included an extrapolation with $\omega=1.61$ fixed.  

Since the simulations were not catered for pairwise intersections, it
may be unfair to directly compare the extrapolations in Table
\ref{tab:Comparison-of-couplings}. Still, due 
to the rapid convergence of $\beta_{c}(N_{t},N_{s})$, our data was
quite well centered around the intersection points except for the
smallest lattices. And knowing what value of $Z_{tw}/Z_{0}$ to concentrate
the efforts around was a big advantage of our method. Anticipating
the location of pairwise intersection points is much more troublesome.
In all, it's clear that the exploitation of the universal number
$F_{tw}(\beta_c) = \ln(1+2^{3/4}) $ gave a significant
boost to the precision of our results.

\begin{table}

\begin{center}
\begin{tabular}{cc|@{\hskip 8pt}cc}
$N_{s}^{'}-N_{s}$ & intersection coupling & $N_{s}$ & $\beta_{c}(4,N_{s})$ from (\ref{eq:linear fit})\\[1pt]
\hline
&&&\\[-11pt]
\hline
&&&\\[-9pt]
12-24 & 6.5539(19) & 24 & 6.53240(14)\\
16-32 & 6.5451(12) & 32 & 6.53459(23)\\
24-48 & 6.53816(37) & 48 & 6.53569(16)\\
32-64 & 6.53706(45) & 64 & 6.53611(19)\\
48-96 & 6.53756(94) & 96 & 6.53648(37)\\[1pt]
\hline
&&&\\[-9pt]
Extrap. & 6.53558(45) &  & 6.53640(10)$^*$ \\[1pt]
\hline
&&&\\[-9pt]
$\chi^{2}$/dof.  & 1.45 &  & 1.62
\end{tabular}
\end{center}

\caption{Comparison of critical couplings obtained for $N_t \!=\! 4$ via
  pairwise intersection (left) and via intersection with the universal
  reference line (right). $^*$This is the extrapolation with all 8
  available $N_s$ values included, as quoted in the $8^\mathrm{th}$
  column of Table \ref{tab:Critical-beta-values}. The
  restriction to the 5 values of $N_s \in (24, 96)$ listed here gives
  6.53641(5), which is consistent but has an unnaturally small error.} 
\label{tab:Comparison-of-couplings}

\end{table}

\subsection{2D Ising model}

\begin{figure}
\leftline{\hskip -.2cm
\includegraphics[width=1.04\columnwidth]{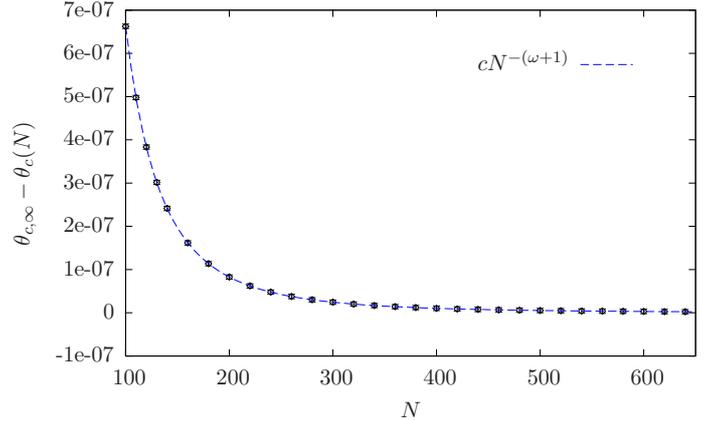} }
\caption{Critical\label{fig:convergenceising} coupling estimate vs $N$ for
the 2D Ising model together with a fit by  (\ref{eq:2DIsingfit})
yielding $\omega = 2.0002$ for $N_\mathrm{min}=100 $, or $\omega = 2.00001$ 
for $N_\mathrm{min} = 400$.}

\end{figure}

For the sake of comparison we repeated our procedure for the square
2D Ising model. Taking the exact solutions for $Z_{ap}$ and $Z_{pp}$
on $N\times N$ lattices from Ref. \cite{1999JPhA...32.4897W}, we used
Mathematica \cite{mathematica} 
to find $\theta = k_{B}T/J$ at the intersection of $Z_{ap}/Z_{pp}$
with the universal value (\ref{eq:universal ratio}). Here $J$ is
the coupling of nearest neighbour spins. In Figure \ref{fig:convergenceising}
we plot the results for $N\in [100,640] $. The error bars are representative
only (the errors are limited only by the working precision and should
be smaller than $10^{-16}$).    
%

Again we fit the data with a FSS ansatz of the form 
\begin{equation}
\theta_{c}(N)=\theta_{c,\infty}-cN^{-(\omega+1)}. \label{eq:2DIsingfit}
\end{equation}
In this case, we hold $\theta_{c,\infty}$ fixed at the known value
of $2/\ln(1+\sqrt{2})$. We obtain for the correction to scaling exponent
$\omega=2 + \delta$ with $\delta\to 0^+$ as we increase the lower
bound $N_\mathrm{min}$ used in the fit, {\it i.e.}, $\omega$ tends
towards 2 from above ($\delta $ starts at around $2\cdot 10^{-4}$ for the
full range of $N$ shown in Fig.~\ref{fig:convergenceising}, and it
falls below $10^{-5}$ at $N_\mathrm{min}$ around 400).     
Since the exponent of the leading irrelevant operator that breaks
rotational invariance is predicted to be exactly 2
\cite{PhysRevE.57.184},  
our result is consistent with the conjecture of Ref.~\cite{Caselle:2001jv}
that the only irrelevant operators that appear in the 2D nearest neighbour
Ising model are those due to the lattice breaking of rotational
symmetry. This may be tested on a triangular 2D lattice where the leading
irrelevant operator to break rotational invariance leads to
$\omega = 4$ instead, while the leading rotationally invariant
operator would give an isotropic correction to scaling with $\omega =
2 $ in either case \cite{Caselle:2001jv}.

On the other hand, our correction exponent for SU(2) is clearly at
odds with $\omega=2$. In this case, it's possible that there exists
an irrelevant operator that is not present in the 2D Ising model 
or the corresponding conformal theory. It may be more likely, however,
that our exponent is really an effective exponent. When there are several
nearby competing exponents it is extremely difficult to extract the 
smallest one from simulations.\footnote{It was noted in
  \cite{Engels:1985tz} that the observed correction to scaling
  exponent of $2+\!1$-dimensional SU(2), 
  $\omega = 1.64 $ in their case or $\omega = 1.61(9) $ in ours,
  agreed well with some predictions for the universality class of the
  2D Ising model which included $ 1.6$ \cite{LeGuillou:1985pg}. At the
  time it was discussed whether such non-integral correction exponents
  could arise in other ferromagentic models of this class, and whether
  the corresponding correction amplitudes happened to vanish
  identically in the exactly solvable pure Ising model, see
  \cite{Barma:1985zz}. This seems to be ruled out by a conformal field
  theory analysis: there is no irrelevant operator with  $\omega < 2 $ in 
  any unitary model of the 2D Ising class, see
  \cite{Calabrese:2000dy,Caselle:2001jv}.}   
   
\subsection{\label{sub:betac vs nt}$\beta_{c}$ vs $N_{t}$}

\begin{figure*}
\includegraphics[width=1\linewidth]{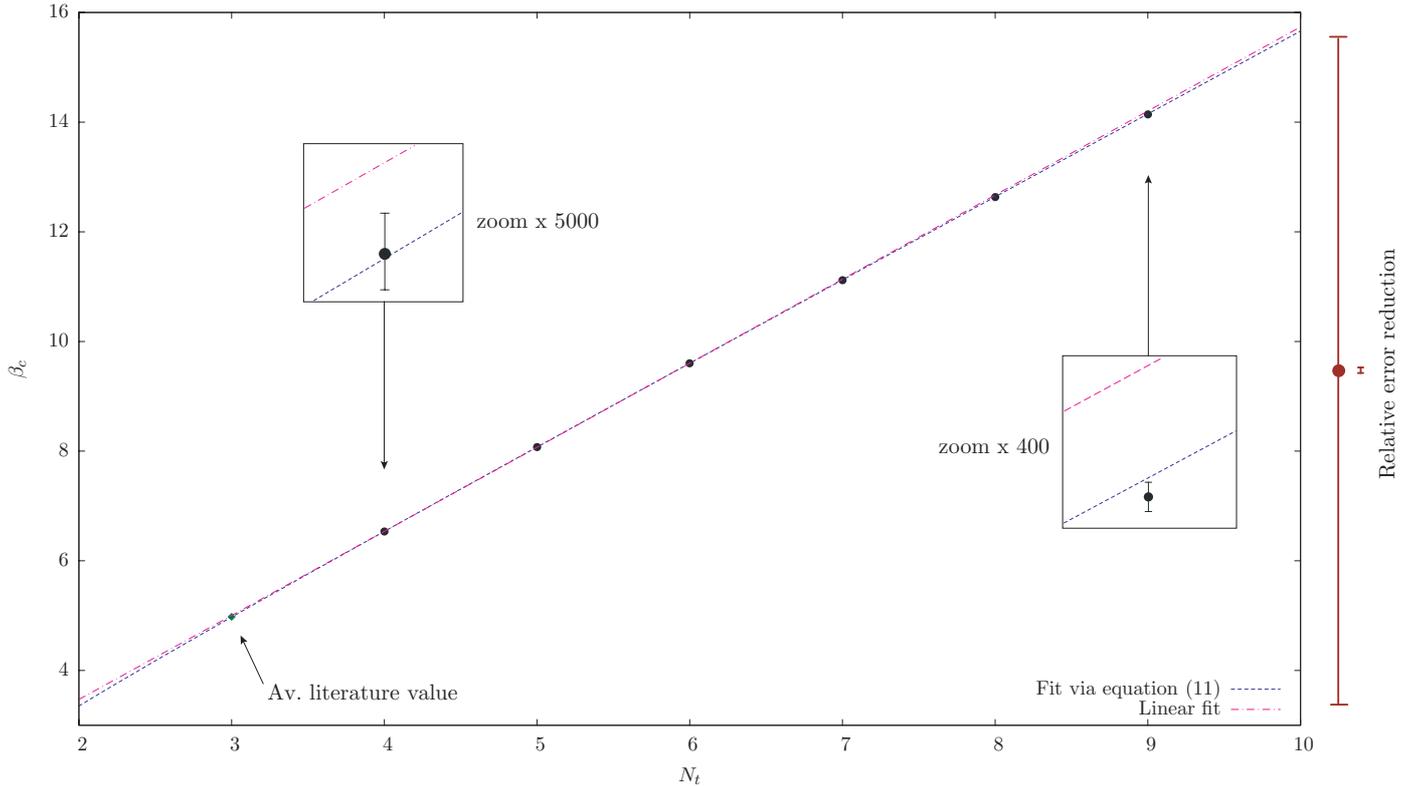}

\caption{$\beta_{c}(N_{t})$ vs $N_{t}$. Note the deviation
from linearity.\label{beta vs nt}}

\end{figure*}

In 2+1 dimensions the coupling $g_{3}^{2}$ has the dimension of mass and
sets the scale for the theory. The bare lattice coupling
is then given by \cite{Engels:1996dz} 
\begin{equation}
\beta=\frac{2N_{c}}{ag_{3B}^{2}}\, , \quad
g_{3B}^{2}=g_{3}^{2}+c_{1}ag_{3}^{4}+c_{2}a^{2}g_{3}^{6}+\dots
\label{eq:lattice coupling}
\end{equation}
 Substituting this expansion into $\beta$ gives
\begin{equation}
\frac{\beta}{2N_{c}} 
  =  \frac{1}{ag_{3}^{2}}-c_{1}-c_{2}ag_{3}^{2}+\dots \, .
\end{equation}
Note now that $T=1/(N_{t}a)$. So at criticality we have
\begin{equation}
\frac{\beta_{c}(N_{t})}{2N_{c}}=\frac{T_{c}}{g_{3}^{2}}\, N_{t}-c_{1}-
c_{2} \, \frac{g_{3}^{2}}{T_{c}} \, \frac{1}{N_{t}}+\dots
\label{eq:crit beta fit}
\end{equation}
The first correction to the bare lattice coupling gives the y-intercept
of \textbf{$\beta_{c}(N_{t})$ }vs $N_{t}$ while the second is a
correction to linearity for small $N_{t}$.

In Fig. \ref{beta vs nt} we plot our results for $\beta_{c}(N_{t})$,
using the fitted values with $\omega$ free. We include also the estimate
$\beta_{c}(3)=4.978(35)$, which is the average of the literature
values 4.943(13) \cite{Liddle:2008kk} and 5.013(15) \cite{Teper:1993gp}. 
For the uncertainty we use their standard error, which is simply half
their difference. The actual error used here has very little influence on
the results.  
The data is fitted with both a straight line and with a function of
the form (\ref{eq:crit beta fit}). To the naked eye, it would appear
that the linear approximation is good all the way down to $N_{t}=3$.
However, our data is precise enough to pick up the deviation from
linearity. In fact, the reduced $\chi^{2}$ of the linear fit is $\sim 70$.
With, the $1/N_{t}$ correction the reduced $\chi^{2}$ drops to 1.5.
Thus, the data is very well described by Eq. (\ref{eq:crit beta fit}).
We obtain the parameterisation
\begin{equation}
\beta_{c}(N_{t})  =
1.5028(21)\, N_{t}+0.705(21)-0.718(49)\, \frac{1}{N_{t}} \, . 
\label{eq:crit beta fit values}
\end{equation}
Consequently, from Eq.~(\ref{eq:crit beta fit}) for
$N_c = 2$,  we
obtain
\begin{equation}
  \label{eq:Tcvg}
     T_c \, = \, 0.3757(5) \; g_3^2 \, .
\end{equation}
This is compatible with the estimate $T_c/g_3^2 \simeq 0.385 $ with an
error of $ \pm 0.010 $ as quoted in Ref.~\cite{Teper:1993gp}. 
Note that it is the $1/N_t$ corrections included in 
(\ref{eq:crit beta fit}) that lead to a somewhat lower value here as
compared to \cite{Teper:1993gp} which is however within their estimate
of the systematic uncertainties. In order to translate our estimate
into units of the zero-temperature string tension $\sigma $ we use the mean
value with standard error of the four $N=2$ values for $\sqrt{\sigma}/g_3^2$
listed in \cite{Bringoltz:2006zg} as pairs of upper and lower bounds
including some systematic uncertainty, which gives
$\sqrt{\sigma}/g_3^2 = 0.3347(5)$. With uncorrelated error propagation,
this together with our estimate (\ref{eq:Tcvg}) then corresponds to
\begin{equation}
  \label{eq:Tcvg}
     T_c/\sqrt{\sigma} \, = \, 1.1225(23) \, ,
\end{equation}
which agrees very well with the corresponding result of
\cite{Liddle:2008kk}, $ T_c/\sqrt{\sigma} \, = \, 1.1224(90) $.   

From Eq.~(\ref{eq:crit beta fit values}), the other constants in
Eq.~(\ref{eq:crit beta fit}) are analogously determined as
\begin{equation}
  \label{eq:consts_1}
  c_1 \, = \, - 0.176(5) \, , \;\; c_2 = 0.0675(5) \, .
\end{equation}
Eq.~(\ref{eq:crit beta fit values}) can be used to obtain accurate
estimates of the critical coupling  at large values of $N_{t}$. 
We can furthermore include the finite-size corrections in 
our 'pseudo-critical coupling' (\ref{eq:beta scaling}) for the
intersection, at finite $N_s$, of the vortex free energy $F_{tw}$ in
(\ref{eq:linear fit}) with the universal value $\ln (1+2^{3/4})$. 
If we assume a form as given in Eq.~(\ref{eq:d_fit_2}), these
corrections can be expressed entirely in terms of the aspect
ratio $A\equiv N_t/N_s$. In particular, the leading corrections for small
$ A $ can then conveniently be combined with the large $N_t $
expansion (\ref{eq:crit beta fit}) as follows,
\begin{eqnarray}
  \label{eq:pseudo_crit_beta_fit}
  \beta_c (N_t,N_s) &=& \Big( 4 \, \frac{T_{c}}{g_{3}^{2}}- d_2 \,
  A^\rho  \Big) \,  N_{t} - 4c_{1} - d_1\, A^{\rho} \\
   && \hskip .2cm  -    
   \Big( 4 c_{2} \, \frac{g_{3}^{2}}{T_{c}} + d_0\, A^{\rho} \Big) 
 \, \frac{1}{N_{t}}+\dots \, , \nonumber
\end{eqnarray}
where we again used $N_c=2$ and $\rho = \omega + 1/\nu$. Since 
$\nu = 1$ in the 2D Ising model, our global fit to the correction to
scaling exponent $\omega $ amounts to $\rho = \omega\!+\!1=2.61(9)$.
If we assume $d_1 = 0$, our fits yield 
\begin{equation}
  \label{eq:d_fit_values_1}
  d_2 = 0.134(4) \, , \;\; d_0 = -0.51(9) \; ,
\end{equation}
and a reduced $\chi^2$ of 2.7. With $d_0 = 0$ on the other hand, 
\begin{equation}
  \label{eq:d_fit_values_1}
  d_2 = 0.155(7) \, , \;\; d_1 = -0.21(4) \; ,
\end{equation}
and a reduced $\chi^2$ of 2.6. The corresponding fits are shown in
Fig.~\ref{fig:D_Nt}. Apart from this systematic uncertainty in the otherwise
elegant parametrisation  (\ref{eq:d_fit_2}) of the finite-size corrections, we
have therefore determined all parameters in
(\ref{eq:pseudo_crit_beta_fit}).

\subsection{\label{sub:betac vs T}$\beta$ vs  temperature}
    
So far, we have studied the critical couplings for different
$N_t$. These were measured from the intersection of the vortex free
energy with the universal value. Since $T = 1/N_ta$, and with the
temperature kept fixed at $T_c$, this means that the corresponding
lattice spacing at criticality scales inversely with $N_t$, {\it i.e.}, for
two lattices with $N_t$ and $N_t'$ we have $a_c'/a_c = N_t/N_t'$, and 
at the given order in $1/N_t$ (in the infinite volume limit),  
\begin{equation}
  \label{eq:beta_c_with_Nt}
  \frac{\beta_c(N_t')}{4} - \frac{\beta_c(N_t)}{4} = \frac{T_c}{g_3^2}
  \, (N_t'-N_t) - c_2 \frac{g_3^2}{T_c} \Big(\frac{1}{N_t'} -
  \frac{1}{N_t} \Big) + \dots \, . 
\end{equation}
Alternatively, we can consider a change of $N_t$ as a change in
temperature at a fixed lattice spacing. In particular, a simulation
done at criticality of the $N_t'$ lattice, with $a_c(N_t') = a(N_t)$,
then corresponds to a simulation at $T = (N_t'/N_t) T_c$ on the $N_t$
lattice, {\it i.e.}, with $\beta \equiv \beta(a)$, we have 
\begin{equation}
  \label{eq:beta_fixed_a}
  \beta_c(N_t') = \beta(1/(N_t'T_c)) = \beta (1/(N_tT))\, .
\end{equation}
Denoting this coupling by $\beta(T,N_t) $, we can  
therefore use Eq.~(\ref{eq:beta_c_with_Nt}) to write down an
equation for the temperature dependence of the lattice coupling near
criticality at a fixed $N_t$,
\begin{equation}
  \label{eq:beta_T_fixed_Nt}
  \frac{\beta(T,N_t)}{4} - \frac{\beta_c(N_t)}{4} = \frac{N_t}{g_3^2}
  \, \big(T-T_c\big) - c_2 \frac{g_3^2}{N_t} \Big(\frac{1}{T} -
  \frac{1}{T_c} \Big) + \dots \, . 
\end{equation}
If we compare simulations at different $N_t$ but with the same aspect ratio
$A = N_t/N_s$ in a finite volume, we can furthermore include the finite-size
corrections here as in (\ref{eq:pseudo_crit_beta_fit}). In terms of the reduced temperature $t = T/T_c-1$, we then have
\begin{align}
  \label{eq:beta_t_fixed_ar}
  \beta(T,N_t,N_s) &=  \beta_c(N_t,N_s)  +  C(N_t,N_s) \, t + \dots \, , \\
   C(N_t,N_s) &=
\Big( 4 \, \frac{T_{c}}{g_{3}^{2}}- d_2 \,
  A^\rho  \Big) \,  N_{t} + \Big( 4 c_{2} \, \frac{g_{3}^{2}}{T_{c}} +
  d_0\, A^{\rho} \Big)   \, \frac{1}{N_{t}} \, , \nonumber 
\end{align}
where $\beta_c(N_t,N_s) $ is the pseudo-critical
coupling in Eq.~(\ref{eq:pseudo_crit_beta_fit}),
 and $\beta(T,N_t,N_s) $ refers to
the temperature dependence of the coupling near this $\beta_c$ at
fixed $N_t$ and $N_s$. The terms neglected are either subleading in
the aspect ratio $A$, or of order $1/N_t^2$, or they are of higher
order in the reduced temperature. Note that criticality in a finite volume is a
fuzzy concept, of course. As before, it is here defined by $\beta =
\beta_c$ at $t=0$ from the intersection of the vortex free energy with the
universal value, {\it c.f.}, Eq.~(\ref{eq:linear fit}).

Note that the linear relationship between lattice coupling $\beta$ and
temperature in Eq.~(\ref{eq:beta_t_fixed_ar}), which is of the form
\begin{equation}
  \label{eq:T_vs_beta}
  T/T_c = 1 + (\beta-\beta_c)/C \, ,
\end{equation}
can be used to control the temperature at fixed $N_t$ and $N_s$ 
by adjusting the lattice coupling. The corresponding procedure for
SU(2) in 3+1 dimensions, where $\beta$ depends logarithmically on
temperature, was used in \cite{deForcrand:2001nd}.
We will exploit this to perform a detailed finite-size scaling
analysis of the vortex free energy $F_{tw}$ in 2+1 dimensional SU(2)
with high precision in a future paper. This will
include scaling of data from various $N_t$ lattices.

\section{Summary}

\vspace{-.1cm}

We have shown that very accurate determinations of the critical
couplings for the SU(2) deconfinement transition in 2+1 dimensions are
possible from intersecting vortex free energies with the known universal
value from the 2D square Ising model at criticality. This allowed us
to determine the critical coupling for lattices with up to $N_t=9$
sites in the Euclidean time direction. Its $N_t$ dependence could be
determined with an accuracy sufficient to require significant $1/N_t$
corrections to the linear behaviour of the continuum limit. As a
result, the slope of this behaviour given by $T_c/g_3^2$ might have
been somewhat overestimated in the past. In particular, we observe that
rather large $N_t$ may be required to reach the asymptotic continuum
behaviour, even though it should be approached much more rapidly here
than in 3+1 dimensions.

\vspace{-.3cm}

\section*{Acknowledgements}

\vspace{-.2cm}

Communications with Martin Hasenbusch are greatly acknowledged. We
thank him for helpful comments, for pointing out
Ref. \cite{Caselle:1995wn} to us and for an enlightening unpublished
note on the corrections to scaling of the Binder cumulant of the 2D
Ising model. Our simulations were performed mainly on the high
performance computing facilities provided by eResearch South
Australia. This work was supported by the Helmholtz International
Center for FAIR within the LOEWE program of the State of Hesse.

\end{document}